\documentclass[english,showpacs,floatfix,12pt]{revtex4}
\usepackage[dvips]{graphicx}
\usepackage{longtable}
\usepackage{amsmath,amssymb}
\usepackage{dcolumn}
\usepackage{latexsym}
\usepackage{babel}
\usepackage[latin1]{inputenc}
\usepackage{color}
\usepackage[colorlinks]{hyperref}
\def\al{\alpha}
\def\kp{\kappa}

\def\pa{\partial}
\def\vf{\varphi}

\def\om{\omega}
\def\Ga{\Gamma}
\def\ga{\gamma}
\def\be{\beta}

\def\dl{\delta}

\def\th{\theta}

\def\wt{\widetilde}
\def\nn{\nonumber}
\def\diag{\mbox {diag}}

\def\l{\left}
\def\r{\right}

\begin{document}
\title{\Large Light-Cone Coordinate System in General Relativity}
\date{5th December 2017}%\today}%19th May 2017}%5th November 2017}%18th November 2017}%
\author{Ying-Qiu Gu}
\email{yqgu@fudan.edu.cn} \affiliation{School of Mathematical
Science, Fudan University, Shanghai 200433, China}


\begin{abstract}
If there is a null gradient field in 1+3 dimensional space-time, we
can set up a kind of light-cone coordinate system in the space-time.
In such coordinate system, the metric takes a simple form, which is
much helpful for simplifying and solving the Einstein's field
equation. This light-cone coordinate system has wonderful properties
and has been widely used in astrophysics to calculate parameters. In
this paper, we give a detailed discussion for the structure of
space-time with light-cone coordinate system. We derive the
conditions for existence of such coordinate system, and show how to
construct the light-cone coordinate system from usual ones, then
explain their geometrical and physical meanings by examples.

\end{abstract}

\maketitle

\section{Introduction}
\setcounter{equation}{0} A good choice of coordinate system for the
space-time is important to discuss the property of space-time and
solve the Einstein's equation. The usual choices are the Gaussian
normal coordinates system and the harmonic coordinates
system\cite{1}. These coordinate systems bring about some
convenience for theoretical analysis. However, such coordinate
systems are not any helpful to solve the Einstein's field equation.
The conventional method to get the exact solution of Einstein's
equation is usually based on the symmetry of the space-time. Many
well-known solutions such as the Friedmann-Lemaitre-Robertson-Walker
metric, Bianchi universe, Lemaitre-de Sitter universe, as well as
Schwarzschild metric and Kerr metric, Taub-NUT
solution\cite{2,3,rot,mac,Zgu1}, are all related with some special
symmetry of the space-time.

In this paper, we study the structure of light-cone coordinate
system(LCS). In such LCS, some partial differentials in Einstein's
tensor $G_{\mu\nu}$ can decouple into ordinary derivatives. This
property was very helpful to solve the exact vacuum solutions of
Einstein's field equation\cite{gu2} and simplify the dynamics of an
evolving star\cite{gu1}. This coordinate system can be constructed
from a set of null geodesics. There were some coordinate systems
related to light-cone used in the previous researches. In Minkowski
space-time we have `light-cone coordinate'.  In Schwarzschild
space-time, we have `Eddington-Finklestein coordinates'. The
Newman-Penrose formalism is also based on null tetrad\cite{NP1}.
However it is a little different from LCS and the practical
calculation in this formalism is not easy.

In recent years, the geodesic light-cone coordinates(GLC) is
introduced  to derive explicit expressions for averaging the
redshift to luminosity-distance relation in a generic inhomogeneous
universe\cite{r8r}. It is also related to the light-cone coordinates
and shares many common properties of LCS. The advantages and
wonderful properties of GLC were recognized by more and more
researchers. Some pedagogical introduction to GLC and brief review
on its applications are provided in \cite{rvw,g19,rvw2}. The GLC is
exploited to perform light cone averages in a perturbed
Friedmann-Lemaitre-Robertson-Walker space-time, in order to
determine the effect of inhomogeneities on the distance-redshift
relation \cite{g7, g8,g9,g10,g11,g13,g19}, and therefore on the
interpretation of the Hubble diagram \cite{g12,g13,g14}.  GLC is
also applied to gravitational lensing in general\cite{g15,g16}, to
galaxy number counts\cite{g17}, and to the propagation of
ultra-relativistic particles\cite{g18}. The presence of additional
degrees of freedom in the GLC was considered later in
\cite{r16r,r17r}. In \cite{r18r}, the correct prediction of GLC
approach in the conformal Newtonian gauge is compared with other
approaches.  After the correction suggested in \cite{r16r}, the GLC
approach has been successfully used to calculate the expressions of
the light-cone observables up to second order in perturbation theory
in the Poisson gauge\cite{g11,g17,g16,r19r}. The consistency of GLC
approach with the previous results in an inhomogeneous universe is
considered in \cite{r24r}.

In this paper, we discuss the structure of a space-time with LCS in
details, and establish the relationship between LCS with ordinary
coordinate system. It is a modification of the early version
arXiv:0708.2962v1. The conditions for a LCS are derived, and the
differential equations to construct a LCS from usual coordinate
system are obtained. Some typical examples to set up a LCS are
given.

\section{construction of light-cone coordinate system}
\setcounter{equation}{0} Under some conditions, the metric in a LCS
has the following simple form
\begin{eqnarray} g_{\mu\nu}=\left( \begin {array}{cccc} u &v&p&q\\ \noalign{\medskip}v&0&0&0\\
\noalign{\medskip}p&0&-{a} &0\\\noalign{\medskip}q&0&0&-{b}
\end {array} \right).
\label{2} \end{eqnarray} At first, we give some general analysis for
the coordinate transformation to get this canonical metric
(\ref{2}). The line element of space-time is given by
\begin{eqnarray}
ds^2=g_{\mu\nu} d\xi ^\mu d\xi ^\nu=d Z^+ G dZ,\qquad
dZ=(d\xi^0,d\xi^1,d\xi^2,d\xi^3)^+, \label{dsx}
\end{eqnarray} where $G=(g_{\mu\nu})$ is matrix form of metric, and
index `$+$' stands for transpose. In this paper, we use Greek
characters such as $\mu,\nu\in \{0,1,2,3\}$ to stand for
4-dimensional indices, and Latin characters $k,l\in \{1,2,3\}$ for
spatial indices. Making transformation $\xi ^\mu=\xi ^\mu(x^\nu)$
and denoting
\begin{eqnarray}
Y_\mu=\l(\frac {\pa \xi ^0}{\pa x^\mu},\frac {\pa \xi ^1}{\pa
x^\mu},\frac {\pa \xi ^2}{\pa x^\mu},\frac {\pa \xi ^3}{\pa
x^\mu}\r)^+,\qquad J=(Y_0,Y_1,Y_2,Y_3), \label{dfz}
\end{eqnarray}
where $J$ is the Jacobian matrix of transformation, we get
\begin{eqnarray}
ds^2=dX^+ J^+ G J dX,\qquad dX=(dx^0,dx^1,dx^2,dx^3)^+. \label{dsy}
\end{eqnarray}
If $x^\mu$ forms light-cone coordinate system, by (\ref{2}) and
(\ref{dsy}) we have
\begin{eqnarray}
(Y_1,Y_2,Y_3)^+ G Y_1=0,\label{z1} \\ Y_2^+ G Y_3=0. \label{z23}
\end{eqnarray}
Let $Y_0^+ G Y_1=K\ne 0$, then by (\ref{z1}) we get
\begin{eqnarray}
J^+ G Y_1=K(1,0,0,0)^+, \qquad Y_1 = K G^{-1}J^*(1,0,0,0)^+,
\label{z0}
\end{eqnarray}
where $J^*=(J^{-1})^+$.

Noticing the unidirectionality of time, we assume
\begin{eqnarray}
 x^0=T(\xi ^\mu),\qquad \pa_0 T>0. \label{y0}
\end{eqnarray}
In component form, (\ref{z0}) becomes
\begin{eqnarray}
\frac {\pa \xi ^\mu}{\pa x^1}=K g^{\mu\nu} \frac{\pa x^0} {\pa \xi
^\nu}= Kg^{\mu\nu}\pa_\nu T. \label{z1c}
\end{eqnarray}
Since $x^0$ and $x^1$ are independent variables in new coordinate
system $x^\mu$, we have
\begin{eqnarray}
0=\frac {\pa x^0}{\pa x^1}=\pa_\mu T\frac{\pa \xi ^\mu} {\pa x^1}=K
g^{\mu\nu} \pa_\mu T \pa_\nu T. \label{null}
\end{eqnarray}
This means the time coordinate transformation $T(\xi ^\mu)$ is a
null gradient field. (\ref{null}) is a necessary condition for LCS.

Let $V_\mu=\pa_\mu T$, if $V^0\ne 0$, by (\ref{z1c}) as $\mu=0$, and
then using (\ref{null}) we have
\begin{eqnarray}
K=\frac 1 {V^0}\frac {\pa \xi ^0}{\pa x^1}=- \frac {
V_k}{V_0V^0}\frac {\pa \xi ^k}{\pa x^1}. \label{KK}
\end{eqnarray}
Substituting (\ref{KK}) into (\ref{z1c}) we get a homogeneous linear
equation for $\pa_1 \xi ^k$
\begin{eqnarray}
\l(\dl^k_n V^0V_0 +V^k V_n \r) \frac{\pa \xi ^n}{\pa x^1}=0.
\label{lnfn}
\end{eqnarray}
The determinant of the coefficient matrix is given by
\begin{eqnarray}
\det\l(\dl^k_n V^0V_0 +V^k V_n \r) =\l(V_0 V^0\r)^2 V_\mu V^\mu=0.
\label{lnfn1}
\end{eqnarray}
The solution reads
\begin{eqnarray}
\frac \pa {\pa x^1} \xi ^k=f(\xi ^\mu)V^k. \label{xfn}
\end{eqnarray}
(\ref{xfn}) is also in the form of (\ref{z1c}), but in (\ref{xfn})
$f\ne 0$ is an arbitrary function which can be selected according to
requirement. Solving $\xi ^0$ from (\ref{y0}), we have
$\xi^0=t(x^0,\xi ^k)$. Substituting it into (\ref{xfn}),  we get an
ordinary differential equation system of $\xi ^k(x^1)$ for any given
$f$. We have a unique solution for initial problem
\begin{eqnarray}
\xi ^k=F^k(x^0,x^1,X^l), \label{xsl}
\end{eqnarray}
where $X^k$ is the initial values of $\xi ^k$.  Making any
differentiable and invertible transformation
\begin{eqnarray}
X^k=X^k(x^0,x^2,x^3), \label{xsl1}
\end{eqnarray}
substituting it into (\ref{xsl}), and then substituting the results
into $\xi ^0=t(x^0,\xi ^k)$, we get a transformation
\begin{eqnarray}
\xi ^\mu=\xi ^\mu(x^0,x^1,x^2,x^3). \label{xsl0}
\end{eqnarray}

In new coordinate system $(x^0,x^1,x^2,x^3)$, equation (\ref{z1})
holds. However, we have still two problems. The first is if the null
gradient field $g^{\mu\nu}\pa_\mu T\pa_\nu T=0$ has nontrivial
solution $\pa_\mu T \not \equiv 0$, and under what condition we have
nontrivial solution. The second is under what conditions (\ref{z23})
holds.  In what follows, by means of light cone we discuss the
problem in detail. The analysis shows, the existence of nontrivial
solution of $T(\xi^\mu)$ is equivalent to the existence of a series
of global null geodesics $\xi ^\mu(\tau)$ in the space-time, and we
have
\begin{eqnarray}
\pa^\mu T \propto \frac {d \xi ^\mu} {d\tau}. \label{null1}
\end{eqnarray}

Directly solving the null gradient field $T(\xi^\mu)$ from
(\ref{null}) is difficult. However, $T(\xi^\mu)$ can be equivalently
derived from null geodesics and the LCS can be constructed as
follows.

 {\bf Theorem 1}. {\em There is a LCS in a space-time, i.e., the metric can be transformed into the following form
\begin{eqnarray} g_{\mu\nu}=\left( \begin {array}{cccc} u &v&p&q\\ \noalign{\medskip}v&0&0&0\\
\noalign{\medskip}p&0&-{a} & s\\\noalign{\medskip}q&0&s&-{b}
\end {array} \right),
\label{nort} \end{eqnarray} if and only if there is a null vector
field
 $V^\mu$ in the space-time $V_{\mu}V^\mu=0$, and the 1-form
\begin{eqnarray}\om=g_{\mu\nu}V^\mu d \xi^\nu \label{omg}\end{eqnarray}
is integrable.}

{\bf Proof.} For necessary part, since $(t,z,x,y)$ is the light-cone
coordinate system, solving the null geodesics along the $z$ axis in
the space-time with metric (\ref{nort}), we get
\begin{eqnarray}
\frac {d^2 z}{d\tau^2}=-\frac {\pa_z v} v \left(\frac {d
z}{d\tau}\right)^2 , \quad \frac {dt}{d\tau}=\frac {dx}{d\tau}=\frac
{dy}{d\tau}= 0.
\end{eqnarray}
The solution of the null vector field is given by
\begin{eqnarray}
V^\mu\equiv \frac d{d\tau} x^\mu =\left(0,~\frac \kp  v,~
0,~0\right),
\end{eqnarray}
where $\kp $ is a constant. The 1-form (\ref{omg}) becomes
\begin{eqnarray}\om=g_{tz}V^z d t=\kp  dt, \label{omg1}\end{eqnarray}
which is an exact differential form.

For the sufficient part, assume $d\xi^0$ to be time-like. Define
\begin{eqnarray} dt= \wt K g_{\mu\nu}V^\mu d \xi^\nu,\label{3} \end{eqnarray}
where $\wt K$ is a factor to make the 1-form (\ref{3}) become an
exact differential form, it satisfies
\begin{eqnarray}
\frac {\pa t}{\pa{\xi^0}}=\wt K g_{\mu 0}V^\mu >0.\quad (\forall
\xi^0).
\end{eqnarray} Then we have a regular coordinate transformation for $t$
\begin{eqnarray}
t=T(\xi^\mu),\qquad V_\mu =  \wt K^{-1} \pa_\mu T. \label{4}
\end{eqnarray}

Along any null geodesic with tangent vector $V^\mu= \frac {d
}{d\tau}\xi^\mu(\tau)$, where $\tau$ is the parameter of the
geodesic, we have
\begin{eqnarray} dt(\tau)=\pa_\mu T \frac {d \xi^\mu(\tau)}{d\tau}d \tau
= \wt K V_\mu V^\mu d\tau=0.\label{2.3.1}
\end{eqnarray} So for any given constant $t_0$, the hypersurface
$T(\xi^\mu)=t_0$ is a propagating light wave front orthogonal to
$V^\mu$. That is to say, the geometrical meaning of hypersurface
$T(\xi^\mu)=t_0$ is a light wave front scanning the space.

Now we construct the coordinate $z=z(\xi^\mu)$, which describes the
distance of the light wave front $T(\xi^\mu)=t_0$ moving through
\begin{eqnarray} dz=\frac {\pa z}{\pa \xi^\mu} d\xi^\mu.\label{4.2} \end{eqnarray}
Taking the trajectories of the null geodesic, namely the light rays,
as the $z$ axes, then along these $z$ axes we have $V^\mu= \frac {d
}{d\tau}\xi^\mu(\tau)$, Substituting it into (\ref{4.2}) we get
\begin{eqnarray} V^\mu \pa_\mu z= \frac {dz}{d\tau}\equiv K,\label{4.4} \end{eqnarray}
where $K\ne 0$ is a smooth function to be determined, which acts as
the scale of $z$ axis. If $K<0$, make an inversion transformation
$\wt z=-z$, we get $V^\mu \pa_\mu\wt z= |K|$. So not lose
generality, we always assume $K>0$.

If we choose $K$, such that the 2-d surface
$t(\xi^\mu)|_{z=const.}=t_0$ is always a fixed light wave front. We
denote it by $S(t_0,z)$. The initial surface is $S_0=S(t_0,z_0)$,
where $(t_0,z_0)$ are given constants. By the definition, $S$ is
orthogonal to null vector $V^\mu$, that is,  $S$ is always
orthogonal to the light rays---z axes. Solving (\ref{4.4}) with
boundary condition $z|_{{S_0}}=z_0$ on surface ${S_0}$, we obtain
the coordinate transformation $z=z(\xi^\mu)$. The moving distance of
the propagating light wave front $S{(t_0,z_0)}\to S{(t_0,z)}$
defines the new coordinate $z$.

For the 2-d surface $S(t_0,z_0)$, not loss generality, we can assume
the parameter coordinates $(x,y)$ are orthogonal grid. Otherwise, we
can take the 2 principal curves of the surface as coordinate grid of
$(x,y)$ to get orthogonal coordinates. If we set each null geodesic
with unique parameter coordinate $(x,y)$, then the coordinates
$(x,y)$ become global coordinates.  The metric in new coordinate
system $(t, z, x, y)$ takes the following form
\begin{eqnarray} g_{\mu\nu}=\left( \begin {array}{cccc} u &v&p&q\\ \noalign{\medskip}v&-w &0&0\\
\noalign{\medskip}p&0&-{a} &s\\\noalign{\medskip}q&0&s&-{b}
\end {array} \right).
\label{10} \end{eqnarray}

For light travels along the $z$ lines, we have $dx=dy=0$, and the
line element becomes
\begin{eqnarray}
0=ds^2=u dt^2-w dz^2+2vdt dz.\label{11} \end{eqnarray} By the
definition of $t$ in (\ref{2.3.1}), we have $dt=0$ for the same
propagating light wave front $S{(t_0,z_0)}\to S{(t_0,z)}$. In this
case $dz\ne 0$, so we get $w=0$ from (\ref{11}). Considering the
arbitrary of $(t, z, x, y)$, we have $w\equiv 0$, and then we get
the metric (\ref{nort}). The proof is finished.

The selection of $K$ in (\ref{4.4}) is quite arbitrary.  For
convenience of solving (\ref{4.4}), we can usually take $K=1$ or
$K=z$ or some factors of vector $V^\mu$ to make the equation
simpler.

{\bf Theorem 2}. {\em Assuming in LCS $(t,z,\wt x,\wt y)$ the metric
takes the form (\ref{nort}), then we have

$1^\circ$ Let $A=s a^{-1}, B=s b^{-1}$, If $\pa_z A=0$ or $\pa_z
B=0$, there exists a regular coordinate transformation, such that
(\ref{nort}) can be converted into canonical form (\ref{2}).

$2^\circ$ In the general case with $\pa_z a\ne 0$, $\pa_z b\ne0$ and
$\pa_z s\ne0$, metric (\ref{nort}) can be converted into canonical
form (\ref{2}) if and only if there exist $(A,B)$ independent of
$z$, such that $s$ satisfies
\begin{eqnarray} s=\frac {A a+B b}{1+AB},\qquad \pa_z A=\pa_z B=0,
\label{fcnd} \end{eqnarray} and the following partial differential
equation system for transformation $\wt x(t,x,y)$, $\wt y(t,x,y)$
has regular solution,}
\begin{eqnarray}
{\pa_y\wt x}=A{\pa_y \wt y},\qquad {\pa_x\wt y}=B {\pa_x \wt x}.
\label{AB}
\end{eqnarray}

{\bf Proof.} In the case $\pa_z B=0$, substituting transformation
$\wt x=x$, $\wt y=\wt y(t,x,y)$ into the line element $ds^2$, we get
metric (\ref{2}) by
\begin{eqnarray} \frac {\pa \wt y}{\pa x}=B(t, x,\wt y).  \label{ycnd} \end{eqnarray}
Taking $t$ as an independent parameter, (\ref{ycnd}) becomes an
ordinary differential equation for $\wt y(x)$. Solving it we get a
unique solution for initial value problem $\wt y= f(t,x,y_0)$.
Making any regular transformation $y_0\leftrightarrow y$, or
concretely $y_0=f(t,y)$, we get the total transformation $\wt y=\wt
y(t,x,y)$. Similarly we can check the case $\pa_z A =0$ by
transformation $\wt x=\wt x(t,x,y)$, $\wt y=y$.

For the case in $2^\circ$, since $z$ axes are the light rays which
have been selected, the coordinates transformation  $(\wt x,\wt
y)\leftrightarrow (x,y)$ must be independent of $z$. Under some
transformation $\wt x =\wt x (t,x,y)$, $\wt y=\wt y(t,x,y)$, the
metric should be converted into (\ref{2}). By straightforward
calculation, we find $s$ should take the form of (\ref{fcnd}), and
the solution of (\ref{AB}) gives the transformation to convert
metric (\ref{nort}) into (\ref{2}). The proof is finished.

The condition  (\ref{fcnd}) is similar to a conformal condition for
the 2-d surface $S(t,z)$ for different $z$. Since $S(t_0,z_0)\to
S(t_0,z)$ is an equidistant translation of grid $(x,y)$ along
geodesics, (\ref{fcnd}) is a natural requirement for the space-time
with LCS. Whether (\ref{fcnd}) can be proved by geometry or derived
from vacuum Einstein's equation $G_{\mu\nu}=0$ is still a problem.
If the metric satisfies the conditions in Theorem 2, all space-like
coordinates can be orthogonalized, and then the spatial coordinates
$(z,x,y)$ form an global orthogonal coordinate grid. The new metric
(\ref{nort}) becomes the canonical form (\ref{2}).

{\bf Theorem 3}. {\em Assuming the coordinate system $(t,z,x,y)$ is
LCS and the metric
 takes the canonical form (\ref{2}), under the following coordinate transformation,
\begin{eqnarray}
t=f_0(t'),~~z=f_1(t',z'),~~x=f_2(t',x'),~~y=f_3(t',y'),
\label{trans}
\end{eqnarray}
where $f_\mu$ are any given smooth functions, the metric also takes
the canonical form (\ref{2}) in new coordinate system
$(t',z',x',y')$.}

Theorem 3 can be directly checked.

From the above proof, we find that the new coordinate system
$(t,z,x,y)$ is induced from a global null geodesic series, so it is
worthy of the name `{\bf light-cone coordinate system}'. In such
LCS, the structure of the space-time becomes simpler, and the exact
solutions to the Einstein's field equation can be more easily
obtained\cite{gu1,gu2}. For an evolving star with spherical
symmetry, in LCS the Einstein's field equation can be reduced to
some ordinary differential equations\cite{gu1}.

The GLC introduced in \cite{r8r,rvw,rvw2,g19} has a little
difference from (\ref{nort}). In GLC the signature of metric is
chosen as $(-,+,+,+)$ and the line element is given by
\begin{eqnarray}
ds^2_{\rm GLC}=-2\Upsilon d\tau d w +\Upsilon^2 dw^2
+\ga_{kl}(d\th^k-U^k dw)(d\th^l-U^l dw),\quad (k,l)\in \{1,2\}.
\label{glc}
\end{eqnarray}
The LCS version is inclined to theoretical discussion but GLC is
inclined to applications in astrophysics. The basic properties of
LCS and GLC are quite similar and can refer to each other. However,
one constraint of coordinate condition in GLC is given by
$g_{ww}=\Upsilon ^2+\ga_{kl} U^k U^l$. The specific relations
between two kinds light-cone coordinate system need to be clarified
in details.

{\bf Theorem 4}. {\em If there is a null gradient field $V_\mu
V^\mu=0$ in space-time $\xi^\mu$, $(t,z,x,y)$ is a LCS.  Then the
coordinate transformation functions $\xi^\mu\leftrightarrow
(t,z,x,y)$ satisfy the following linear partial differential
equations
\begin{eqnarray} V^\mu \pa_\mu (t,z,x,y)=(0,f,0,0),
\label{12} \end{eqnarray} in which $f(t,z)> 0$ is any given function
with suitable smoothness}.

{\bf Proof.} For the function $t(\xi^\mu)$, by $\pa_\mu t\propto
V_\mu$ we have
\begin{eqnarray} V^\mu \pa_\mu t(\xi^\nu)\propto V^\mu V_\mu=0.
\label{eq1} \end{eqnarray} So $t=t(\xi^\mu)$ satisfies (\ref{12}).

The coordinate $z(\xi^\mu)$ is defined by (\ref{4.4}), so it also
satisfies (\ref{12}).

For the coordinate function $x$, along $z$ axes we have
\begin{eqnarray} 0=dx=\pa_\mu x d\xi^\mu=\wt K^{-1} V^\mu \pa_\mu x d\tau,
\label{13} \end{eqnarray} that is $x=x(\xi^\mu)$ satisfies
(\ref{12}). Similarly, we can check $y=y(\xi^\mu)$ also satisfies
(\ref{12}). The proof is finished.

(\ref{12}) forms the basic differential equation system to determine
the light-cone coordinate system $(t,z,x,y)$. The above derivation
shows the physical and geometrical meanings of LCS and the
corresponding metric (\ref{nort}). It also provides a method to
solve null gradient field equation $\pa_\mu T\pa^\mu T=0$.

\section{examples and applications}
\setcounter{equation}{0}

At first, we take some simple cases in Minkowski space-time as
examples to show  concepts of LCS. We have line element
\begin{eqnarray}
ds^2=d\xi_\mu d\xi^\mu=
(d\xi^0)^2-(d\xi^1)^2-(d\xi^2)^2-(d\xi^3)^2.\label{m1}
\end{eqnarray}
The simplest case corresponds to the plan wave moving along
$z=\xi^1$, we have
\begin{eqnarray}
(t,z,x,y)=(\xi^0-\xi^1,\xi^1,\xi^2,\xi^3),\label{m2}  \\
 ds^2=dt^2+2dt dz -d x^2-dy^2.\label{m3}
\end{eqnarray}
$dt=0$ means $\xi^0=\xi^1+t_0$, which stands for a propagating wave
front $S(t_0,z)$. $dt=dz=0$ corresponds to a fixed wave front
$S(t_0,z_0)=\{\xi^\mu |\xi^0=t_0+z_0, \xi^1=z_0\}$.
\begin{eqnarray}
V_\mu=K\pa_\mu t=K(1,-1,0,0),\quad V^\mu=K(1,1,0,0),\quad V_\mu
V^\mu=0.\label{m3*}
\end{eqnarray}
Let $V_0V^0=1$ we get $K=1$.

 The second case corresponds to  cylinder wave  moving along
$\rho$, we have
\begin{eqnarray}
&(t,\rho,\phi,z)=(\xi^0-\rho,\sqrt{(\xi^1)^2+(\xi^2)^2}, \arctan ({\xi^1}/{\xi^2}),\xi^3),&\label{m4}  \\
 &ds^2=dt^2+2dt d\rho -\rho^2 d \phi^2-dz^2.&\label{m5}
\end{eqnarray}
The third case corresponds to the spherical wave moving along $r$,
we have
\begin{eqnarray}
t=\xi^0-r,\qquad  ds^2=dt^2+2dt d r -r^2( d \th^2+\sin^2\th
d\vf^2).\label{m6}
\end{eqnarray}

In what follows, we take Schwarzschild space-time and Kerr-like one
as examples to explain the geometrical meaning of LCS and show how
to construct the LCS.

For the Schwarzschild metric
\begin{eqnarray}
g_{\mu\nu}=\diag \left[ 1 -\frac {2m} r,~ -\left(1 -\frac {2m}
r\right)^{-1},~ -r^2, ~-r^2\sin^2\th\right], \quad (r>2m)
\label{3.1}
\end{eqnarray}
with the coordinate system $(t, r, \th, \vf)$, the radial null
geodesic satisfies
\begin{eqnarray}
g_{00}\dot t^2-g_{11}\dot r^2=\left(1 -\frac {2m} r\right)\dot
t^2-\left(1 -\frac {2m} r\right)^{-1}\dot r^2=0,\label{3.2}
\end{eqnarray}
where $\dot t =\frac {dt}{d\tau}$ and $\dot r =\frac {dr}{d\tau}$.
Taking the null vector orthogonal to the 2-d surface $(\th,\phi)$ as
follows
\begin{eqnarray}
V^\mu=\left( \left(1 -\frac {2m} r\right)^{-1},~\pm
1,~0,~0\right),\label{3.3}
\end{eqnarray}
it is easy to check that the corresponding 1-form (\ref{omg}) is an
exact differential form. The initial light wave front $S(t_0,r_0)$
is simply a sphere in the domain $r>2m$. $V^r=1$ corresponds to the
outward light rays and $V^r=-1$ corresponds to the inward light
rays. In what follows we only calculate the case $V^r=1$.

By (\ref{3}), we get the coordinate function $\wt t $
\begin{eqnarray}
\wt t = \int g_{\mu\nu} V^\mu d\xi^\nu= t-r-2m\ln(r-2m)+t_0.
\label{3.4}
\end{eqnarray}
By (\ref{12}), for $x,y$ we have
\begin{eqnarray}
V^\mu\pa_\mu F=\left(1 -\frac {2m} r\right)^{-1}\pa_t F +\pa_r
F=0.\label{3.5}
\end{eqnarray}
The general solution is given by $F=H(\wt t ,\th,\phi)$, where
$H(\wt t ,\th,\phi)$ is arbitrary smooth function. By boundary
condition, we get
\begin{eqnarray}
x=\th,\qquad y=\phi. \label{3.7}
\end{eqnarray}

By (\ref{4.4}), we have
\begin{eqnarray}
V^\mu\pa_\mu z=\left(1 -\frac {2m} r\right)^{-1}\pa_t z +\pa_r
z=\left(1 -\frac {2m} r\right)^{-1} f.\label{3.8}
\end{eqnarray}
For (\ref{3.8}), we get typical solutions independent of $(x, y)$
\begin{eqnarray}
z=\left \{ \begin{array}{ll}
  r+2m\ln(r-2m)+Z(\wt t ),  & {\mbox{if}} \quad  f=1,\\
  Z(\wt t )(r-2m)^{2m}e^r,  & {\mbox{if}} \quad  f=z,\\
  Z(\wt t )+r, & {\mbox{if}} \quad  f=\left(1 -\frac {2m} r\right),
\end{array} \right. \label{3.9}
\end{eqnarray}
where $Z(\wt t )$ is an arbitrary function of $\wt t $, we can set
$Z=z_0$ according to Theorem 3. In fact, we can choose any given
monotone increasing function $z(r)$ in this case,
\begin{eqnarray}f=\left(1 -\frac {2m} r\right) z'(r).\end{eqnarray}
So the option of $f>0$ is quite arbitrary. The Eddington-Finklestein
coordinates are similar to these coordinate system.

In the case of the metric generated by rotating source similar to
the Kerr ones\cite{rot}, we cannot generally construct a null vector
field $V^\mu$ satisfying the integrable 1-form (\ref{omg}), so the
corresponding metric cannot be generally converted into the
canonical form (\ref{2}). Now we examine the  following metric in
the coordinate system $(t, r, \th, \phi)$,
\begin{eqnarray} g_{\mu\nu}=\left( \begin {array}{cccc} u^2 &0&0&uw\\ \noalign{\medskip}0&-a&0&0\\
\noalign{\medskip}0&0&-b&0\\\noalign{\medskip}uw&0&0&w^2-v
\end {array} \right),
\label{3.10} \end{eqnarray} where $u,v,w,a,b$ are smooth functions
of $(r,\th)$, but independent of $(t,\phi)$.  For speed
\begin{eqnarray}
V^\mu=\left(\dot t,~\dot r,~\dot \th,~\dot \phi\right),\label{3.11}
\end{eqnarray}
after some arrangement, the geodesic equation $ \dot
V^\al=-\Ga^\al_{\mu\nu}V^\mu V^\nu$ becomes
\begin{eqnarray}
\frac d{d\tau} \dot t &=&- \frac 1{uv}\left((2v-w^2)\frac {d
u}{d\tau}+uw\frac {d w}{d\tau}\right)\dot t\nn\\
&&-\frac 1 {u^2v}\left(w(v-w^2)\frac {d u}{d\tau} +u(v+w^2)\frac {d
w}{d\tau}-uw\frac {d v}{d\tau}\right)\dot \phi,\label{3.12}\\
\frac d{d\tau} \dot \phi &=&- \frac 1{v}\left(w\frac {d
u}{d\tau}-u\frac {d w}{d\tau}\right)\dot t-\frac 1
{uv}\left(w^2\frac {d u}{d\tau} -uw\frac {d w}{d\tau}+u\frac {d
v}{d\tau}\right)\dot \phi,\label{3.13}
\end{eqnarray}
\begin{eqnarray}
\frac {d } {d\tau} (a \dot r^2) &=&-\left(2u\pa_r u\dot t^2+
(2w\pa_r u+2u\pa_r w)\dot t\dot
\phi+(2w\pa_r w-\pa_r v)\dot \phi^2\right)\dot r\nn\\
&&+\pa_r b \dot r\dot \th^2 - \pa_\th a \dot r^2\dot \th,\label{3.14}\\
\frac {d } {d\tau} (b \dot \th^2) &=&-\left(2u\pa_\th u\dot t^2+
(2w\pa_\th u+2u\pa_\th w)\dot t\dot
\phi+(2w\pa_\th w-\pa_\th v)\dot \phi^2\right)\dot \th \nn\\
&&-\pa_r b \dot r\dot \th^2 + \pa_\th a \dot r^2\dot
\th.\label{3.15}
\end{eqnarray}
(\ref{3.12}) and (\ref{3.13}) are integrable due to the two Killing
vectors $(\pa_t,\pa_\phi)$. The first integrals of (\ref{3.12}) and
(\ref{3.13}) are given by
\begin{eqnarray}
\dot t=-m\frac{w^2-v}{vu^2}-n\frac w{uv},\qquad\dot\phi=m\frac
w{uv}+n\frac 1 v, \label{3.16}
\end{eqnarray}
where $m,n$ are constants. Substituting (\ref{3.16}) into the line
element equation, we have an equation for null geodesic
\begin{eqnarray}
a\dot r^2+b\dot \th^2=\frac {m^2}{u^2}-\frac {(nu+mw)^2}{u^2v}.
\label{3.17}
\end{eqnarray}

By (\ref{3.10}) and (\ref{3.16}), the covariant speed becomes
\begin{eqnarray}
V_\mu=g_{\mu\nu}V^\nu=(m,~-a\dot r,~-b\dot \th,~ -n). \label{3.18}
\end{eqnarray}
$V_t$ and $V_\phi$ are constants related with the Killing vectors
$(\pa_t,\pa_\phi)$.

According to Theorem 4, the metric can be converted into
(\ref{nort}) if and only if there exists a function
$T(t,r,\th,\phi)$ such that  $V_\mu=\pa_\mu T$ is a null vector.
Then by (\ref{3.18}), we have
\begin{eqnarray}
\pa_tT&=&m,\qquad~~~ \pa_\phi T=-n, \label{3.18.1}\\
\pa_r T&=& -a\dot r,\qquad \pa_\th T= -b\dot \th.\label{3.18.2}
\end{eqnarray}
Solving (\ref{3.18.1}) we get
\begin{eqnarray}
T=mt-k\th-n\phi-h(r,\th),\label{3.19}
\end{eqnarray}
where $k$ is a constant. $k\th$ is split from $h(r,\th)$ for
simplicity of following calculation. Substituting (\ref{3.19}) into
(\ref{3.18.2}) we get
\begin{eqnarray}
\dot r =\frac 1 a \pa_r h,\qquad \dot \th =\frac 1 b (k+\pa_\th
h).\label{3.20}
\end{eqnarray}
By (\ref{3.19}) we find $m$ is the scale of time, so we set $m=1$.
Substituting (\ref{3.20}) into (\ref{3.14}) and (\ref{3.15}), we get
\begin{eqnarray}
(\pa_r h)^2 = a \left(\frac {1}{u^2}-\frac {(nu+w)^2}{u^2v}-\frac
{k^2} b\right),\quad \pa_\th h=0,\quad k\in  \{
  0,  1\}. \label{3.21}
\end{eqnarray}
By (\ref{3.21}), we find $h=h(r)$.

(\ref{3.21}) includes many cases of space-time with LCS. We only
discuss the case $k=0$ in normal spherical coordinate system. By
(\ref{3.21}) and $k=0$, we get
\begin{eqnarray}
a= \frac {u^2vh'(r)^2}{v-(nu+w)^2}. \label{3.22}
\end{eqnarray}
(\ref{3.22}) is a necessary condition that the metric (\ref{3.10})
can be converted into (\ref{nort}) in the case $k=0$.

Comparing the Kerr metric in the Boyer-Lindquist form with
(\ref{3.10})\cite{2,mac,rot}, we obtain
\begin{eqnarray}
u^2 &=& {\frac {r^2+\al^2\cos^2\th-2mr}{r^2+\al^2\cos^2 \th }}, \label{3.23}\\
v &=& \frac {2\al^2 mr(r^2+\al^2\cos^2\th+6mr)\sin^4
\th}{(r^2+\al^2\cos^2 \th -2mr)(r^2+\al^2\cos^2 \th)}+(r^2+\al^2)\sin^2 \th , \label{3.24}\\
w &=& \frac{4\al
mr\sin^2 \th }{\sqrt{(r^2+\al^2\cos^2 \th -2mr)(r^2+\al^2\cos^2 \th )}}, \label{3.25}\\
a &=& \frac{r^2+\al^2\cos^2 \th }{r^2-2mr+\al^2},\qquad b~=~
r^2+\al^2\cos^2 \th , \label{3.26}
\end{eqnarray}
where $m$ is the mass of a star, and $\al$ is a constant
proportional to the angular momentum. Substituting
(\ref{3.23})$\sim$(\ref{3.26}) into (\ref{3.22}), we find it
contradicts $\pa_\th h =0$, so the Kerr metric cannot be converted
into (\ref{nort}). Or equivalently, we cannot construct a global
light-cone coordinate system in the Kerr's space-time.

Now we transform the metric (\ref{3.10}) with (\ref{3.22}) into the
canonical form (\ref{2}). For (\ref{3.22}), we make transformation
$\wt r=h(r)$, then we remove the function $h(r)$ from the metric in
the new system $(t,\wt r,\th,\phi)$. This process is equivalent to
setting $h(r)=r$.

Substituting $h=r$ and $k=0$ into (\ref{3.19}), we get the new time
coordinate $T=t-n\phi-r$. By symmetry of the boundary condition, we
should have
\begin{eqnarray}
n=0,\qquad T=t-r. \label{3.27}
\end{eqnarray}

By (\ref{3.20}) and $k=0$, the covariant speed $V^\mu$ defined in
(\ref{3.11}) becomes
\begin{eqnarray}
V^\mu=\left(a^{-1}, a^{-1},~0, ~\frac {w}{uv}\right),\quad a =\frac
{u^2v} {v-w^2}. \label{3.28}
\end{eqnarray}
For general functions $(u,v,w)$, (\ref{12}) cannot be solved
explicitly. If we set the scale function in (\ref{12}) as $f=V^r=
a^{-1}$, we can solve the new coordinate
\begin{eqnarray}
z=r+Z(\wt t ,\th,\Phi), \label{3.29}
\end{eqnarray}
where $Z$ is an arbitrary function, we set $Z=0$ for simplicity.
\begin{eqnarray}
\Phi=\phi-\int\frac{u w}{v-w^2}dr. \label{3.30}
\end{eqnarray}

Solving other equations in (\ref{12}), we get $F=F(\wt t
,\th,\Phi)$. We can choose any two independent functions
\begin{eqnarray}
x=X(\wt t ,\th,\Phi),\qquad y=Y(\wt t ,\th,\Phi), \label{3.32}
\end{eqnarray}
as the new coordinates. In the new coordinate system $(\wt t,
z,x,y)$ defined by (\ref{3.27}), (\ref{3.29}) and (\ref{3.32}), the
metric (\ref{3.10}) becomes $(\wt g_{\mu\nu}) = J^* (g_{\al\be})
J^{-1}$.  By calculation, we find that
\begin{eqnarray}
\wt g_{zz}=\wt g_{zx}=\wt g_{zy}=0. \label{3.35}
\end{eqnarray} However, we have $\wt g_{xy}\ne 0$  in general case.
if
\begin{eqnarray}
\frac{u w}{v-w^2}=\frac d {dr} \vf(r), \label{3.36}
\end{eqnarray}
where $\vf(r)$ is any smooth function independent of $\th$, we can
get $\wt g_{xy}= 0$. In this case, we have
\begin{eqnarray}
(\wt t,z,x,y) =(t-r,r,\th,\phi-\vf), \label{3.37}
\end{eqnarray}
and the new metric becomes canonical form.

\section{conclusion}
\setcounter{equation}{0}

The above discussion shows that we can set up a LCS if and only if
there is a null gradient field in the space-time. In such coordinate
system, the metric takes a wonderful canonical form (\ref{nort}),
and the Einstein's field equation becomes simpler\cite{gu2,gu1}.
This coordinate system might be also helpful to understand the
propagation of the gravitational wave. However, LCS has some
limitations in application, because it holds only in average or
approximate sense in usual cases. Another kind of special coordinate
system with unique realistic time is given in \cite{ncs}. The
physical requirement for space-time with LCS is that, there is a
light at some point or on some 2-d surface in the space-time, and
its wave front is stable enough to act as coordinates. Therefore,
the feature of light-cone coordinate system can be summarized in a
powerful word: Beacon the whole world by one light.

\acknowledgments{The author is grateful to his supervisor Prof.
Ta-Tsien Li for his directing and encouragement. I'd like to thank
Prof. G. Veneziano and Prof. A. Marciano for enlightening
discussions}

\end{document}